\documentclass[aps,twocolumn,amsmath,amssymb,floatfix,showpacs,
groupedaddress]{revtex4}

\usepackage{bm,amsmath,amssymb,amsfonts}

\usepackage[dvips]{graphicx}

\usepackage{color}
\definecolor{brown}{rgb}{0.70,0.30,0.05}
\begin{document}

\title{\textcolor{blue}{Exotic electron states and tunable magneto-transport 
in a fractal Aharonov-Bohm interferometer}} 

\author{Atanu Nandy}
\email{atanunandy1989@gmail.com}

\author{Biplab Pal}
\email{biplabpal@klyuniv.ac.in}

\author{Arunava Chakrabarti}
\email{arunava\_chakrabarti@yahoo.co.in}

\affiliation{Department of Physics, University of Kalyani, Kalyani,
West Bengal-741235, India}

\begin{abstract}
A Sierpinski gasket fractal network model is studied in respect of 
its electronic spectrum and magneto-transport when each `arm' of 
the gasket is replaced by a diamond shaped Aharonov-Bohm interferometer, 
threaded by a uniform magnetic flux. 
Within the framework of a tight binding model for non-interacting, 
spinless electrons and a real space renormalization group method we 
unravel a class of {\it extended} and {\it localized} electronic states. 
In particular, we demonstrate the existence of {\it extreme localization} of 
electronic states at a special finite set of energy eigenvalues, 
and an infinite set of energy eigenvalues where the localization gets 
`delayed' in space ({\it staggered localization}). These eigenstates 
exhibit a multitude of {\it localization areas}. The two terminal 
transmission coefficient and its dependence on the magnetic flux threading each 
basic Aharonov-Bohm interferometer is studied in details. Sharp {\it switch on}-
{\it switch off} effects that can be tuned by controlling the flux from outside, 
are discussed. Our results are analytically exact. 
\end{abstract}


\maketitle
\section{Introduction}
\label{intro}
When an electron goes round a closed loop that traps a magnetic flux 
$\Phi$, its wave function gains a phase $\phi=2 \pi \Phi/\Phi_{0}$, where, 
$\Phi_{0}=hc/e$ is the fundamental flux quantum. This simple sentence is at 
the heart of the path-breaking Aharonov-Bohm (AB) 
effect~\cite{aharonov,washburn,markus1,markus2,markus3} 
that has led to an extensive research in the so called 
{\it AB interferometry} which dominated the physics, both experiments 
and theory, in mesoscopic dimensions over the past couple of 
decades~\cite{yacoby1,yacoby2,casse,kubala,kim,kobayashi,
aharony1,aharony2,aharony3,benjamin,aldea,forster,aharony4,
camino,aharony5,kubo1,urban,kubo2}. 

While recent experiments by Yamamoto {\it et al.}~\cite{yamamoto} has inspired  
more experiments on quantum transport in AB interferometers~\cite{aharony6}, 
the earlier theoretical model studies have also played an instrumental role in 
understanding the basic features of electronic states and coherent transport in 
quantum network models in the mesoscopic or nano-dimensional 
systems~\cite{casse,kubala,kim,kobayashi,aharony1,aharony2,
aharony3,benjamin,aldea,forster,aharony4,camino,aharony5,kubo1,urban,kubo2}. 
The present day advanced growth, fabrication and lithographic techniques 
have opened up the possibility of artificial, tailor made lattice structures 
using quantum dots (QD) or Bose-Einstein condensates (BEC) with tremendous 
potential for application in device technologies. Needless to say, this has 
stimulated a considerable volume of theoretical research even in model 
quantum networks with a complex topological 
character~\cite{andrade,cardoso,oliviera}. 

\begin{figure}[ht]
\centering
\includegraphics[clip,width=8cm,angle=0]{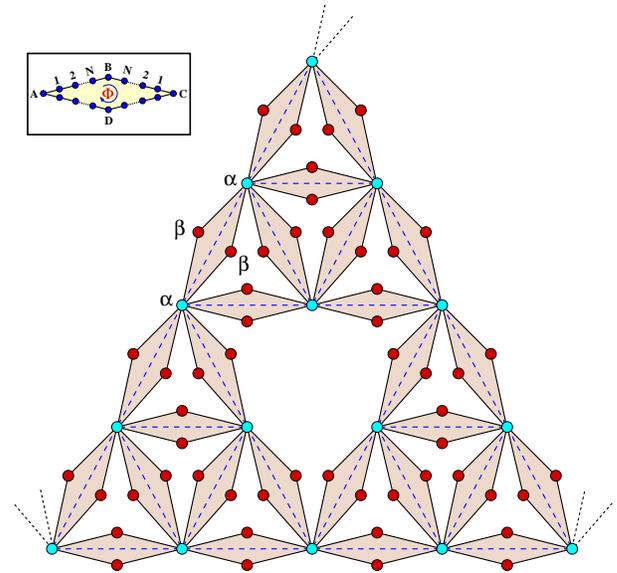}
\caption{(Color online) Schematic view of a part of an infinite fractal Aharonov-Bohm (AB) 
interferometer formed by arranging diamond shaped closed loops on a 
Sierpinski gasket (SPG) matrix. In the inset we have shown a basic plaquette where 
each arm of the diamond loop contains $N$ number of sites in between two vertices.  
A uniform magnetic flux $\Phi$ pierces every closed diamond shaped loop.} 
\label{network}
\end{figure}

In this communication, inspired by the present scenario of theory and 
experiments in AB interferometry, we examine the spectral and transport 
properties of a quantum network in which diamond shaped AB interferometers
are arranged in a Sierpinski gasket (SPG) geometry (Fig.~\ref{network}). 
The SPG geometry is already an extensively studied example of a finitely 
ramified fractal network ~\cite{domany,rammal,banavar,wang}, exhibiting exotic 
electronic spectrum. Apart from it, it has 
also been experimentally exploited to observe dimensionality crossover in 
superconducting wire networks~\cite{gordon}, as well as considered for 
modeling photonic wave guide networks~\cite{li,cooley}. 
We revisit it in the shape of an AB interferometric arrangement of 
model nanorings.

In our work, each elementary interferometer is pierced by a uniform magnetic 
field applied perpendicular to the plane of the diamond plaquette, and 
traps a flux $\Phi$.
Such diamond shaped interferometers have previously been studied as the minimal 
models of bipartite structures having nodes with different coordination numbers, 
and representing a family of itinerant geometrically frustrated electronic systems 
~\cite{vollhardt,lopes1,lopes2}. Other studies include the problem of 
electron localization in the presence of spin-orbit interaction~\cite{bercioux}, 
a flux-induced semiconducting behavior~\cite{sil}, 
quantum level engineering for AB cages~\cite{movilla} or, as models of 
spin filters~\cite{aharony7,aharony8}. Placed in a motif of a recursively grown, 
self-similar, hierarchical pattern of the SPG, such a {\it fractal} AB 
interferometer gives us an opportunity to critically examine the interplay 
of quantum interference and lattice topology in the light of recent developments 
in the subject. 

As a first step in this direction, we examine the spectral properties of 
non-interacting, spinless electrons described in a tight binding formalism. 
Using an exact real space renormalization group (RSRG) decimation scheme we 
show that, the spectrum of such an AB interferometer consists of an infinity 
of extended and localized eigenfunctions induced by the topology of the nanocluster. 
Localized states can belong to at least a couple different categories, and 
can be made to occupy any part of the energy spectrum by tuning the out of plane 
magnetic field. For one class of localized eigenstates, the onset of localization 
can be `delayed' (in space) at will, by extracting the corresponding energy at 
any desired length scale. These are called the {\it staggered} localized states, and 
can be placed at any desired part of the energy spectrum by controlling the 
magnetic flux. This latter control over position of the eigenstates of course, is 
true for all kinds of states.

In addition, the magnetic flux can be used to 
control the width of the gaps in the spectrum, and for special value of flux 
$\Phi=\Phi_{0}/2$, the entire spectrum is shown to collapse into {\it extremely 
localized} eigenstates, a fact known as the AB-caging and pioneered by Vidal {\it et 
al.}~\cite{vidal1,vidal2,vidal3,vidal4}.

In what follows we describe our findings. In section~\ref{sec2} we describe 
the Hamiltonian and the RSRG scheme. Section~\ref{sec3} contains a discussion 
of the density of states and the nature of the eigenfunctions, and in 
section~\ref{sec4} we present the transport calculations. Finally, in 
section~\ref{conclu} we draw our conclusions. 
\section{The beginning}
\label{sec2}
We refer to Fig.~\ref{network}. A basic diamond AB interferometer is shown 
in the inset. Each arm of the diamond contains $N$ number of single level 
QD's ~\cite{kubala} between the vertices, so that the total number of dots 
contained in adiamond is $4N+4$. A magnetic flux $\Phi$ pierces each plaquette, 
and we restrict electron hopping only along the arm of a diamond. 
Spinless, non-interacting electrons are described by the Hamiltonian,  
\begin{equation}
{\bm H} = \sum_{i} \epsilon_{i} c_{i}^{\dagger} c_{i} + \sum_{\langle ij 
\rangle} \left[ t_{ij} e^{i\phi_{ij}}c_{i}^{\dagger} c_{j} + h.c. \right] 
\label{hamilton}
\end{equation}
$\epsilon_{i}$ is the on-site potential at each QD-location, which we assume 
to be constant. That is, $\epsilon_i=\epsilon$ at every vertex $i$. The nearest 
neighbor hopping integral $t_{ij}$ is associated with a typical AB-phase 
$\phi_{ij}=(2\pi/\Phi_{0})\int_{i}^{j} \vec{A} \cdot d\vec{r}$, $\vec{A}$ being the 
vector potential. This phase factor breaks the time reversal symmetry of electron-
hopping along the arms of the interferometer.

Using the set of difference equations satisfied by the electrons, viz., 
\begin{equation}
(E-\epsilon_{i}) \psi_{i} = \sum_{j}t_{ij}e^{i\phi_{ij}} \psi_{j}
\label{difference}
\end{equation}
we first renormalize each diamond plaquette shown in the inset 
of Fig.~\ref{network} into an elementary diamond having just four sites. The {\it 
renormalized} plaquettes are then joined recursively to generate a SPG 
geometry, as shown. Two kinds of sites are generated for an infinite geometry. 
These sites are marked by the cyan dots ($\alpha$) and red dots ($\beta$). 
The coordination numbers for the vertices $\alpha$ and $\beta$ are eight and two 
respectively. On renormalization of the elementary plaquettes (inset), the 
effective on-site potentials at these two sites are given by, 
\begin{equation}
\begin{aligned}
&\epsilon_{\alpha} =  \epsilon + 8t \dfrac{U_{N-1}(x)}{U_N(x)}\\
&\epsilon_{\beta} =  \epsilon + 2t \dfrac{U_{N-1}(x)}{U_N(x)} \\
\end{aligned}
\label{epsrecur}
\end{equation}
the inter-vertex hopping integral acquires a phase, and is given by,
\begin{equation}
t_{F} = t \dfrac{e^{i\phi(N+1)}}{U_{N}(x)}
\label{trecur}
\end{equation}
in the {\it forward} direction and, the {\it backward} hopping is 
$t_{B}=t_{F}^{*}$. Here, $\phi=2\pi\Phi/(N+1)\Phi_{0}$, 
$U_{N}(x)$ is the $N$-th order Chebyshev polynomial of second kind,
and $x=(E-\epsilon)/2t$. 
The familiar SPG network of triangles is easily restored 
from Fig.~\ref{network} by joining the $\alpha$-sites (cyan dots), 
and `eliminating' the $\beta$-sites (red dots).
This leads to the usual triangular SPG geometry (shown by blue dotted lines 
in Fig.~\ref{network}) with only one kind of `bulk' sites having an 
effective on-site potential,
\begin{equation}
\tilde{\epsilon} = \epsilon_{\alpha} + \dfrac{8 t^2}{(E-\epsilon_{\beta})}
\label{site}
\end{equation}
and an effective nearest neighbor hopping integral along an arm of the SPG geometry 
(blue dotted lines) given by,
\begin{equation}
\tilde{t} = \dfrac{2t^2}{(E-\epsilon_{\beta})} 
\cos\left(\dfrac{\pi\Phi}{\Phi_{0}}\right)
\label{hop}
\end{equation} 
We shall now exploit Eq.~\eqref{site} and Eq.~\eqref{hop} to extract information 
about the spectrum, eigenstates and transport provided by such a hierarchical 
AB interferometer.
\section{The RSRG scheme and spectral analysis}
\label{sec3}
\begin{figure}[ht]
\centering
\includegraphics[clip,width=8.5cm,angle=0]{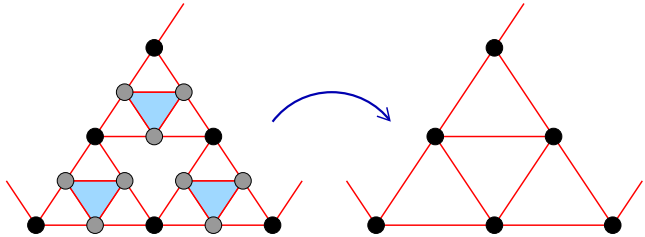}
\caption{(Color online) Decimation renormalization scheme of a basic Sierpinski 
geometry. The grey shaded circles are `eliminated' to generate the next 
re-scaled version.} 
\label{decimation}
\end{figure}

An elementary SPG triangle is shown in Fig.~\ref{decimation}. Both the solid 
black and grey circles have the same on-site potential $\tilde{\epsilon}$.
The first step is to decimate out the grey circles in Fig.~\ref{decimation} 
to obtain 
a renormalized SPG geometry. Only the bulk sites have been shown. 
The renormalized on-site potential and the nearest neighbor hopping integrals 
are given by~\cite{arunava},
\begin{equation}
\begin{aligned}
& \tilde{\epsilon}_{n+1} = \tilde{\epsilon}_{n} + 
\dfrac{4\tilde{t}_{n}^{2} (E-\tilde{\epsilon}_{n})}
{(E-\tilde{\epsilon}_{n})^{2} - 
(E-\tilde{\epsilon}_{n})\tilde{t}_{n} 
- 2\tilde{t}_{n}^{2}} \\
& \tilde{t}_{n+1} = \dfrac{2\tilde{t}_{n}^{3} + 
(E-\tilde{\epsilon}_{n})\tilde{t}_{n}^{2}}
{(E-\tilde{\epsilon}_{n})^{2} - 
(E-\tilde{\epsilon}_{n})\tilde{t}_{n} 
- 2\tilde{t}_{n}^{2}}
\end{aligned} 
\label{recursion}
\end{equation}
In Eq.~\eqref{recursion} $n$ refers to the `stage' of renormalization. 
Therefore, $\tilde{\epsilon}_{0}=\tilde{\epsilon}$, and 
$\tilde{t}_{0}=\tilde{t}$ as given by Eq.~\eqref{site} and Eq.~\eqref{hop} 
respectively.  

\begin{figure}[ht]
\centering
\includegraphics[clip,width=7cm,angle=0]{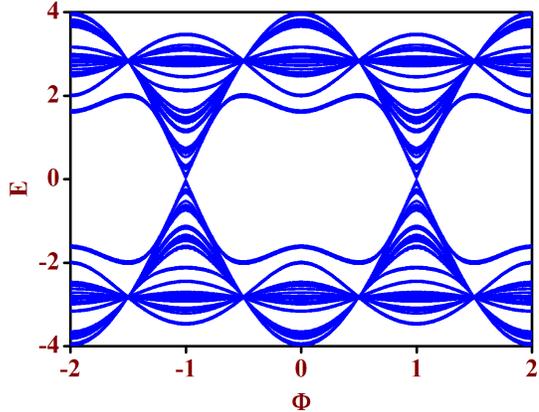}
\caption{(Color online) Flux dependence of energy eigenvalues of an 
infinite $N=0$ AB interferometer based on a Sierpinski gasket geometry.} 
\label{spectrum}
\end{figure}

To get an overall view about the flux dependence of the distribution of 
eigenvalues, we present an energy versus flux diagram for an infinite 
SPG-AB interferometer by evaluating the local Green's function $G_{\alpha}(E)$ at the 
$\alpha$-sites. The simplest picture for the case $N=0$ is shown in 
Fig.~\ref{spectrum}. The elementary diamond AB interferometer (inset of 
Fig.~\ref{network}) now contains just four sites at the vertices. 
The basic building blocks now resemble the $AB_2$ structural unit used by Lopes {\it et 
al.}~\cite{lopes2} in describing their geometrically frustrated systems. 

The spectrum of the SPG-AB interferometer with $N=0$ exhibits a periodicity 
of $2\Phi_{0}$ (Fig.~\ref{spectrum}).
There is a tendency of clustering of the eigenvalues towards the edges of the 
spectrum as is evident from the $E$-$\Phi$ diagram. Multiple band crossings are 
observed and the spectrum exhibits kind of a {\it zero band gap} semiconductor like 
behavior, mimicking a Dirac point as seen in graphene, at $\Phi=\pm \Phi_0$. 
The central gap gets gradually filled up by more eigenstates, and 
the $E$-$\Phi$ contours get more flattened up as $N$ increases. The spectrum at 
$\Phi=\Phi_{0}/2$ consists of only a discrete set of points. This is the case of 
{\it extreme localization}, and happens for any value of $N$, the number of 
eigenvalues in this discrete set depending upon the value of $N$. We shall 
come back to it later. Let us now discuss case by case.
\subsection{Extended States}
It is simple to verify from Eq.~\eqref{recursion} that, if we set 
$E=\tilde{\epsilon}_{n}$, then $\tilde{\epsilon}_{n+1}=\tilde{\epsilon}_{n}$ and, 
$\tilde{t}_{n+2}=-\tilde{t}_{n+1}=\tilde{t}_{n}$ for all subsequent 
values of $n$~\cite{arunava}. 
If a non-zero value of the hopping integral persists at all stages of 
renormalization, it clearly indicates that the amplitudes of the wave function 
have a non-zero overlap between neighboring sites at that scale of length. The 
corresponding energy eigenvalue will thus be of an {\it extended} character.
Depending on the number of QD's in each diamond AB interferometer, the equation 
$E-\tilde{\epsilon}_{n}=0$ will yield a finite number of roots. Each real root 
will then correspond to an eigenvalue of the infinite SPG-AB interferometer for 
which the parameter space ($\tilde{\epsilon}_{n}$,$\tilde{t}_{n}$) 
will exhibit a {\it two cycle} fixed point, 
\begin{equation}
(\tilde{\epsilon}_{n},\tilde{t}_{n}) \rightarrow 
(\tilde{\epsilon}_{n+1},-\tilde{t}_{n+1}) \rightarrow
(\tilde{\epsilon}_{n+2},\tilde{t}_{n+2})
\label{flow}
\end{equation}
The cyclic behavior will set in at a desired value of the RSRG stage $n$, 
at which the roots are extracted. If we make $N \rightarrow \infty$, the roots 
(eigenvalues) will be densely packed in the energy spectrum, forming a 
quasi-continuum. 

For example, with $N=0$, if we set 
$E=\tilde{\epsilon}_{0}=\tilde{\epsilon}=\epsilon + 8t^2/(E-\epsilon)$ 
at the bare length scale ($n=0$), we get two eigenvalues viz., 
$E=\epsilon \pm 2\sqrt{2}t$ which will correspond to extended 
eigenstates in an infinite geometry. The parameter space starts exhibiting a 
two cycle behavior right from the RSRG step $n=1$ onward. These eigenvalues 
are independent of the flux $\Phi$. The eigenvalues obtained from the equation 
$E=\tilde{\epsilon}_{1}$ become functions of the flux $\Phi$. For each of them, 
at any value of the flux $\Phi$, excluding the value $\Phi=\Phi_{0}/2$, the 
hopping integral $\tilde{t}_n$ starts flowing into a two-cycle fixed point for 
$n \ge 2$. 
\begin{figure*}[ht]
\centering
\includegraphics[clip,width=15cm,angle=0]{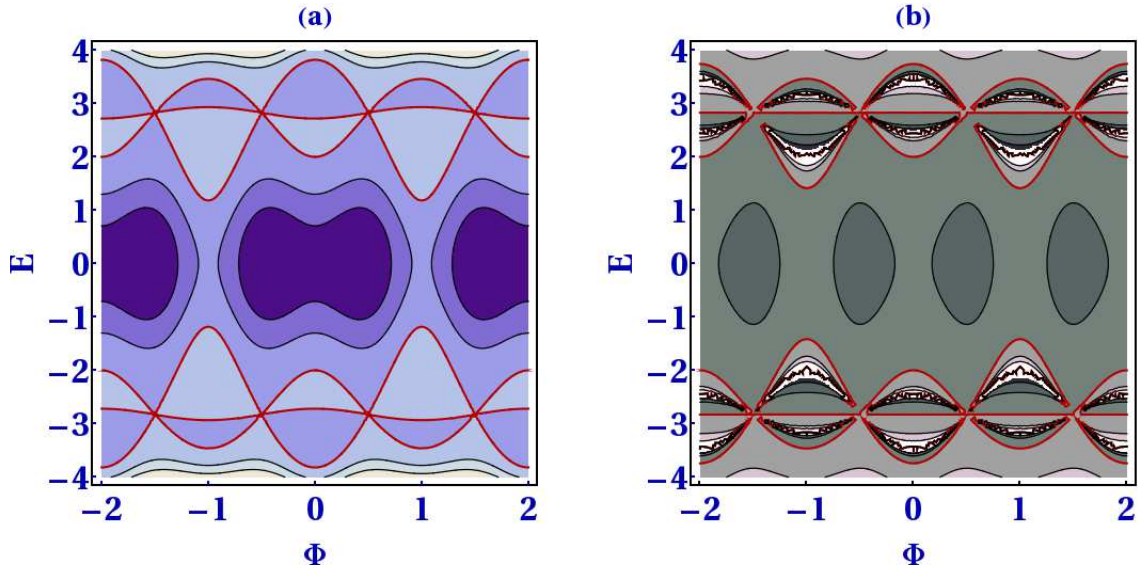}
\caption{(Color online) Contour plots of the functions 
(a) $\mathcal{F}_{1}(E,\Phi) = E-\tilde{\epsilon}_{1}$ and 
(b) $ \mathcal{G}_{1}(E,\Phi) = E-\tilde{\epsilon}_{1} - \tilde{t}_{1}$. 
The red contours in each case indicate a continuous distribution of 
energy and flux for which one can have {\it extended} (in (a)) and the 
{\it staggered localized} eigenstates (in (b)) on a one step 
renormalized lattice.} 
\label{contour}
\end{figure*}
Continuing this argument it is possible to discern a countably infinite 
number of discrete eigenvalues in the limit $n \rightarrow \infty$, and all 
corresponding to extended eigenstates in such a non-translationally invariant 
interferometer arrangement. 
By tuning the magnetic flux appropriately, it is 
possible to `place' such eigenvalues at any desired location in the energy 
spectrum. Such extended eigenstates can, in principle, lie arbitrarily close to 
the neighboring localized states. Thus, by controlling the external magnetic 
field it may be possible to achieve sharp {\it switch on}-{\it switch off} 
effects over arbitrarily small energy intervals.

In Fig.~\ref{contour}(a) we present the contour plot of the function 
$\mathcal{F}_1(E,\Phi) = E - \tilde{\epsilon}_{1}$.
If we progress along the red lines we can have a continuous distribution of 
the energy $E$ and the magnetic flux $\Phi$ for which extended states are 
obtained from a one-step renormalized SPG-AB interferometer. 
The immediate neighborhood of the red contour at any place of the plot is 
largely deviated from the desired result $\mathcal{F}_{1}(E,\Phi)=0$, 
indicating that the extended states are isolated. 
Extended states arising from the other scales of length ($n \ge 2$) 
as solutions of the equation $\mathcal{F}_{n}(E,\Phi)=0$ get densely 
packed in the energy spectrum, but do not form a continuum. 
\subsection{Localization of Eigenfunction}
Absence of any kind of translational invariance makes the appearance of 
localized eigenstates a natural phenomenon. We highlight below some 
interesting variations of these localized states.
\subsubsection{Staggered Localization}
Let's refer back to the set of recursion relations Eq.~\eqref{recursion}. 
The roots of the equation 
$\mathcal{G}_{n}(E,\Phi)= E-\tilde{\epsilon}_{n}-\tilde{t}_{n}=0$ 
yield an interesting result in the sense that, for any value of energy 
extracted out of this equation, the hopping integral $\tilde{t}$ remains 
non-zero upto the $(n+1)$-th stage of renormalization, and starts to decay 
beyond the $(n+1)$-th RSRG step. This implies that for such energy values the 
amplitudes of the corresponding wave function have a non-trivial distribution 
over a cluster of vertices of the SPG-AB inteferometer. The size of the 
cluster increases with increasing value of the RSRG step $n$ from which 
the roots are extracted. 

That such energies indeed belong to the spectrum, 
has been verified by exhaustive calculation of the density of states. As the 
SPG geometry is scaled by a scale factor of three, the size of the 
cluster over which the amplitudes of the wave functions have a non-trivial 
distribution typically gets enlarged in proportion of $3^n$. 
Clusters accommodating the non-trivial distribution of wave functions get 
decoupled from each other (or, remain very weakly coupled at the most) 
beyond a certain scale in such cases.

Since the hopping 
integral $\tilde{t}$ starts flowing to zero beyond the $(n+1)$-th stage, such wave 
functions are indeed localized, but the {\it localization-area} increases 
with increasing value of $n$. In principle, for all such energy eigenvalues 
extracted out of the equation 
$\mathcal{G}_{n}(E,\Phi)= E-\tilde{\epsilon}_n-\tilde{t}_n=0$ 
in the limit $n \rightarrow \infty$, we encounter a countable infinity of 
eigenfunctions which {\it span} the infinitely large interferometer and yet 
retain a localized character. The choice of the RSRG iteration index $n$ thus 
allows one to `delay' (in space) the onset of localization. We name such 
an effect as {\it staggered localization}~\cite{biplab1}.

Contrary to the  case of the extended states, the eigenvalues corresponding 
to the staggered localized states depend on the flux $\Phi$ from the beginning ($n=0$ level), 
and the staggered localization can thus be controlled, at any scale of length, 
to occupy any desired place in the energy spectrum. In Fig.~\ref{contour}(b) 
we show the contour plot of the function $\mathcal{G}_{1}(E,\Phi)$ which bring 
out the location of the energy eigenvalues corresponding to the staggered 
localized states, as marked by the red contours. Such states are again 
isolated, and the number of such states increases with increasing value 
of $n$ (and of course, with the number $N$ of QD's sitting on the arms 
of the elementary diamond interferometer). 
\subsubsection{Localized Edge States}
Existence of localized edge states (LES) in case of an 
ordinary SPG network was pointed out in the pioneering work 
by Domany {\it et al.}~\cite{domany}. We discuss it in terms of the recursion 
relations Eq.~\eqref{recursion}, and unravel the link with the staggered 
localized states discussed above.
\begin{figure}[ht]
\centering
\includegraphics[clip,width=8cm,angle=0]{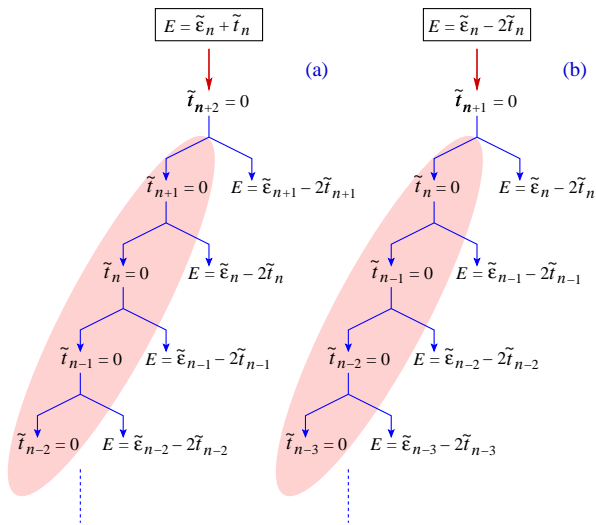}
\caption{(Color online) The genealogical trees for the  
(a) {\it staggered localized} states and (b) {\it edge localized} states of 
the SPG-AB interferometer. In both the figures the highlighted branches of 
the tree correspond to the {\it extreme localization}.}  
\label{tree}
\end{figure}

Let us set $E = \tilde{\epsilon}_{n+1} - 2 \tilde{t}_{n+1}$. 
This immediately makes $\tilde{t}_{n+2}=0$. Then, trivially, hopping 
integrals at all successive stages of renormalization become zero, 
indicating that these eigenvalues correspond to sharply localized states, 
and as argued by Domany {\it et al.}, they represent the so called 
{\it edge states}. It is easy to work out using Eq.~\eqref{recursion}, that 
with the above choice of energy $E$, one arrives at an equation, 
\begin{equation}
\left[(E-\tilde{\epsilon}_{n})^2 - 4\tilde{t}_{n}^{2} \right] 
(E-\tilde{\epsilon}_{n} -\tilde{t}_{n}) = 0
\label{lesroots}
\end{equation}
Clearly, the eigenvalues giving rise to the staggered localization, viz., 
$E=\tilde{\epsilon}_n + \tilde{t}_n$ are all 
embedded in the cluster of eigenvalues obtained from Eq.~\eqref{lesroots}.

Setting $E=\tilde{\epsilon}_n - 2\tilde{t}_n$ yields two branches of solutions.
One gets either $\tilde{t}_n = 0$, or $E = \tilde{\epsilon}_n - 2 \tilde{t}_n$.
Each branch bifurcates into two similar branches as one climbs down the genealogical 
tree of solutions (Fig.~\ref{tree}(b)). If one sticks to the shaded branch only, 
one encounters either a situation where the hopping integral is zero from the 
very beginning. This is a trivial situation, except in one case of the so called 
{\it extreme localization}, as will be discussed immediately after this. The non-shaded 
branch keeps on generating a multitude of the edge-localized eigenstates, in the same 
spirit as discussed by Domany {\it et al.}~\cite{domany}. 
\begin{figure}[ht]
\centering
\includegraphics[clip,width=7cm,angle=0]{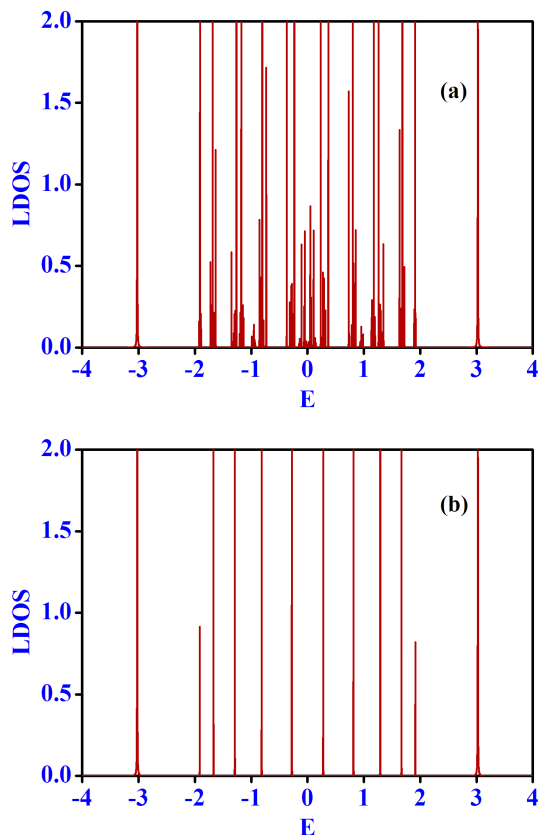}
\caption{(Color online) Local density of states at the bulk ($\alpha$-type) 
sites of the SPG-AB interferometer with $N=5$. The panels exhibit results 
for (a) $\Phi=0$, and (b) for $\Phi=\Phi_{0}/2$. In the latter case, 
the entire spectrum consists of only such {\it extremely localized} states. 
We have selected $\epsilon=0$ and $t=1$, and energy is measured in units of $t$.} 
\label{dos}
\end{figure}
For each root steming from this 
branch at the $n$-th level, the hopping integral drastically drops to zero at the 
$(n+1)$-th stage of renormalization in contrast to the case of staggered localization.

Interestingly, the genealogical tree of the staggered localized eigenstates resulting 
from the equation $E = \tilde{\epsilon}_n + \tilde{t}_n$ yields, as a branch, 
the edge localized states, as is evident from Fig.~\ref{tree}(a). This 
indicates an interesting grouping of the two different kinds of localization, namely the 
staggered and the edge ones. The tuning of the flux can make this grouping 
take place at any desired part of the energy spectrum. The extreme localization 
branch along which $\tilde{t}_n=0$ for any value of $n$, is also obtained as part 
of the solution of the equation $E = \tilde{\epsilon}_n + \tilde{t}_n$.

The subsection of the roots extracted from the equation 
$E-\tilde{\epsilon}_n-2 \tilde{t}_n=0$ is ruled out as exhaustive calculation of the 
density of states indicate that such roots are not part of the energy spectrum.
\subsubsection{Extreme Localization}
From Eq.~\eqref{recursion} it is clear that the eigenvalues corresponding 
to the `extended' eigenstates extracted from the equation $E=\tilde{\epsilon}_{0}$ 
are independent of the magnetic flux threading an elementary diamond interferometer. 
Interestingly, if one sets the magnetic flux $\Phi=\Phi_0/2$, these extended 
states drastically get converted into a set of {\it extremely localized} 
eigenstates in the spirit of the AB cages first pointed out by Vidal {\it et 
al.}~\cite{vidal1,vidal2,vidal3}. The dynamics of the electron is confined within 
clusters of finite size only in all such cases. At half flux quantum, because of the 
presence of the cosine term in Eq.~\eqref{hop} the effective hopping integral connecting 
the vertices of the SPG vanishes confining the amplitudes of the wave function 
in local {\it cages}, the cages being distributed and decoupled from each other 
all over the SPG geometry.

Before ending this section, we discuss the general shape of the 
density of states of an SPG-AB interferometer for $N \ne 0$. 
In Fig.~\ref{dos} we present the local density of states (LDOS) 
at the bulk ($\alpha$, cyan) sites of the AB-SPG interferometer shown in 
Fig.~\ref{network}. Each arm of the basic diamond plaquette now contains $N=5$ 
QD's. The spectrum shown in Fig.~\ref{dos}(a) shows a dense clustering of 
eigenstates which is a mixture of extended, staggered localized, and the 
edge states in the absence of a magnetic flux. The spectrum gradually thins 
out as $\Phi \rightarrow \Phi_{0}/2$, and collapses into a set of 
{\it extremely localized} states only as $\Phi=\Phi_{0}/2$.
\section{Two-terminal transmission characteristics}
\label{sec4}
To get an overview about the nature of the transmittance of such fractal 
network we have computed the two-terminal transmission characteristics 
of a finite generation system. The procedure is standard and is often 
used to evaluate the transmission coefficient of such hierarchical fractal 
structures~\cite{biplab1,biplab2}. We clamp the system in between 
two semi-infinite periodic leads, the so called `source' and the `drain'. The leads 
are characterized by a constant on-site potential ${\mathcal{U}}_{L}$ and 
a constant hopping integral ${\mathcal{T}}_{L}$ in between the atomic sites 
of the leads. The finite sized network sandwiched in between the 
two ordered leads is then successively renormalized to reduce it to an 
effective di-atomic system~\cite{biplab2}. The transmission coefficient of the 
lead-network-lead system is given by a well-known formula~\cite{liu},
\begin{eqnarray}
&T=\dfrac{4\sin^{2}ka}{|\mathcal{A}|^{2}+|\mathcal{B}|^{2}} \\
&\text{with,}\quad \mathcal{A}=[(M_{12}-M_{21})+(M_{11}-M_{22})\cos ka] \nonumber\\
&\text{and}\quad \mathcal{B}=[(M_{11}+M_{22})\sin ka]\nonumber
\end{eqnarray}
where, $M_{ij}$ refer to the dimer-matrix elements, written appropriately 
in terms of the on-site potentials of the final renormalized {\it left} (L) 
and {\it right} (R) atoms $\epsilon_{L}$ and $\epsilon_{R}$ respectively, 
and the renormalized hopping between them. 
$\cos ka=(E-{\mathcal{U}}_{L})/2{\mathcal{T}}_{L}$, 
and $a$ is the lattice constant in the leads which is set equal to unity 
throughout the calculation.

\begin{figure}[ht]
\centering
\includegraphics[clip,width=7cm,angle=0]{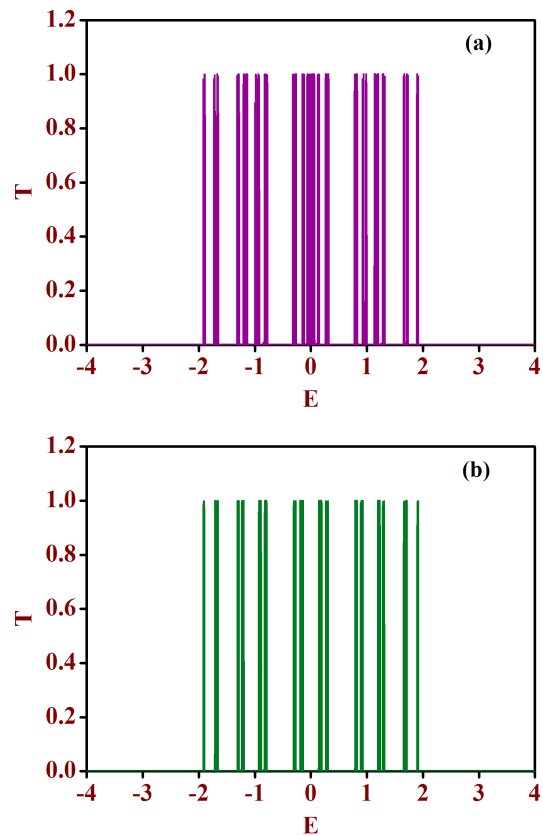}
\caption{(Color online) Transmission characteristics of a $4$-th 
generation SPG-AB interferometer network for (a) $\Phi=0$ 
and (b) $\Phi=\Phi_{0}/4$. The other parameters are $\epsilon=0$, 
$t=1$, $N=5$, and the lead parameters are ${\mathcal{U}}_{L}=0$ and 
${\mathcal{T}}_{L}=2$.} 
\label{trans}
\end{figure}

In Fig.~\ref{trans}, we have shown the variation of the transmission 
coefficient $T$ with the energy of the electron both in presence and in 
absence of external magnetic flux $\Phi$. 
The upper panel shows the 
two-terminal transmission characteristics for a $4$-th generation network 
with $N=5$ in absence of any external magnetic flux. The poor conducting 
behavior of fractal networks in general is reflected in the transmission 
characteristics. There are of course, high, non-zero values of transmission 
coefficient for certain discrete values of energy $E$, which is at par with 
the existence of extended eigenstates in an otherwise fractal character of the 
energy spectrum of the system shown in Fig.~\ref{dos}. For a non-zero value of 
magnetic flux $\Phi=\Phi_{0}/4$ as shown in Fig.~\ref{trans}(b), the number of 
transmitting `zero-widhth windows' of the system gets reduced compared to the 
zero flux case in (a), and finally for $\Phi=\Phi_{0}/2$, i.e., at half flux 
quantum the system becomes completely opaque to an incoming electron.

We have also examined the variation of the transmission coefficient 
with the magnetic flux $\Phi$ for fixed values of electron energy. 
\begin{figure}[ht]
\centering
\includegraphics[clip,width=7cm,angle=0]{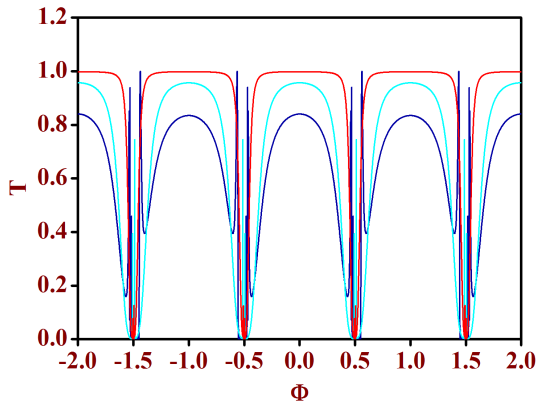}
\caption{(Color online) Aharonov-Bohm oscillations for a $6$-th 
generation SPG-AB interferometer network at different values of 
electron energy, viz., $E=0.27840$ (red line), $E=1.29083$ (cyan line), 
and $E=1.91433$ (blue line). We have chosen $\epsilon=0$, $t=1$, $N=5$, 
and the lead parameters chosen are ${\mathcal{U}}_{L}=0$ and 
${\mathcal{T}}_{L}=2$.   } 
\label{abosc}
\end{figure}
The transmission coefficient exhibits the typical oscillating behavior with 
the variation of the magnetic flux $\Phi$ known as the Aharonov-Bohm (AB) 
oscillation. One interesting feature is to observe that as we increase the 
value of the energy $E$, the AB oscillations get suppressed. In Fig.~\ref{abosc}, 
we have plotted the AB oscillations in the transmittance of a 
sixth generation interferometer for three different values of energy, viz., 
$E=0.27840$ (red line), $E=1.29083$ (cyan line), and $E=1.91433$ (blue line). 
The suppression of AB oscillations is clearly visible.  
\section{Conclusion}
\label{conclu}
In conclusion, we have critically examined the electronic properties of a 
Sierpinski gasket geometry which is formed by arranging diamond shaped 
Aharonov-Bohm interferometers in appropriate order. The energy spectrum 
exhibits a rich structure comprising extended, staggered and edge-localized 
eigenfunctions. The number of such states depend on the number of quantum 
dots in each arm of the elementary diamond interferometer, and can populate 
the energy spectrum as densely as desired by the experimentalists. A constant 
magnetic field can be used to fine tune the locations of all these different 
classes of wave functions to initiate delicate {\it switch on}-{\it switch off} 
effects over arbitrarily small scales of energy. This could be useful from the 
perspective of a switching device that needed to operate over arbitrarily small 
scales of voltage. A magnetic flux equal to half-flux quantum leads to extreme 
localization of the wave function and Aharonov-Bohm cages.
\begin{acknowledgements}
AN acknowledges a research fellowship from UGC, India and BP is 
thankful to DST, India for an INSPIRE fellowship. This work is 
partially supported by DST-PURSE program of University of Kalyani.
\end{acknowledgements}

\end{document}